\documentclass[aps,prl,superscriptaddress,twocolumn]{revtex4-2}

\usepackage{graphicx}
\usepackage{amsmath}
\usepackage{amssymb}
\usepackage{hyperref}
\usepackage[utf8]{inputenc}
\usepackage{mathtools}

\usepackage[english]{babel}
\usepackage{bbm}
\hypersetup{colorlinks=true, linkcolor=blue, citecolor=blue, urlcolor=blue}
\usepackage{xcolor}
\usepackage{braket}
\usepackage[normalem]{ulem}

\newcommand{\re}{\mathrm{Re}}
\newcommand{\im}{\mathrm{Im}}

\newcommand{\fref}[1]{\text{Fig.}~\ref{#1}}

\newcommand{\eref}[1]{\text{Eq.}~\eqref{#1}}

\newcommand{\AffCPH}{Center for Hybrid Quantum Networks (Hy-Q), The Niels Bohr Institute, University~of~Copenhagen,  DK-2100  Copenhagen~{\O}, Denmark}
\begin{document}

\title{Cooperative Sensing with Impurities in a Two-Dimensional Subwavelength Array}

\author{Oliver August Dall'Alba Sandberg}
\email{oliver.sandberg@nbi.ku.dk}
\affiliation{Department of Physics, Harvard University, Cambridge, Massachusetts 02138, USA}
\affiliation{\AffCPH}
\author{Stefan Ostermann}
\affiliation{Department of Physics, Harvard University, Cambridge, Massachusetts 02138, USA}
\author{Susanne F. Yelin}
\affiliation{Department of Physics, Harvard University, Cambridge, Massachusetts 02138, USA}

\begin{abstract}
We propose a versatile quantum sensing protocol based on two dissipatively coupled distant atoms embedded as impurities in a two-dimensional sub-wavelength atomic array.
The array acts as a waveguide for the emitter light, creating cooperative enhancement that allows for more efficient population transfer.
By monitoring the population of one of the impurity atoms, it is possible to detect frequency shifts in the emitters' resonance frequencies. 
We analytically estimate achievable sensitivities as well as the dependence on various system parameters. The proposed protocol is robust against various environmental factors and perturbations, which enhances its applicability in real-world scenarios.
\end{abstract}

\maketitle


\begin{figure}[t]
\includegraphics[width=0.8\linewidth, trim=0cm 0.0cm 0.1cm 0.0cm,clip]{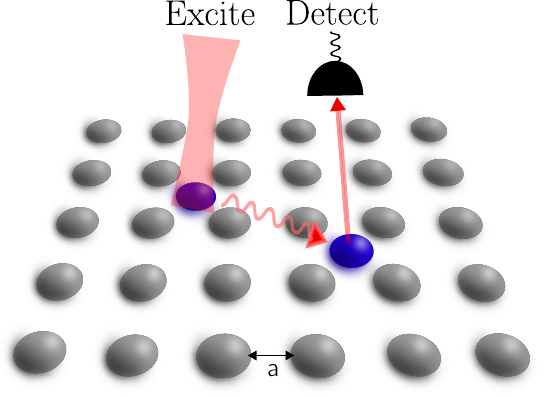}
	\caption{
	A pair of impurities (blue) placed in a sub-wavelength atomic lattice with spacing $a \lesssim \lambda$. 
    In our protocol, we locally excite one atom and detect the population at the other. The lattice atoms couple via optical dipole-dipole interactions and serve as a Markovian bath mediating the interactions and modifying the bare decay rates. We do not invoke the assumption of independent atomic emission, instead the emitted fields of the lattice atoms interfere with each other, giving rise to a coherent collective enhancement to the coupling of the impurities. 
}
	\label{fig:System}
\end{figure}
Efficient light-matter interfaces will be a vital advancement enabling future quantum technologies~\cite{hammerer_quantum_2010}. However, establishing strong coupling between light and matter is challenging due to the small interaction cross-section between matter and single photons, which is in general proportional to the photon's wavelength squared~\cite{gutzler_lightmatter_2021}. 
To overcome this challenge, various platforms which alter the radiative environment by introducing dielectric media near quantum emitters have been developed over the past decade~\cite{mirhosseini_cavity_2019}. 
Besides enhancing the coupling between light and matter on the single photon level, these setups enable long-range couplings between distant emitters and modify emitter decay rates.

Prominent examples of such platforms are quantum emitters, such as quantum dots~\cite{lodahl_interfacing_2015}, nanoparticles~\cite{gonzalez-ballestero_levitodynamics_2021, rieser_tunable_2022}, or atoms coupled to cavities~\cite{rempe_atoms_1993, westmoreland_properties_2019}, nanophotonic waveguides~\cite{vetsch_optical_2010}, photonic band-gap materials~\cite{hood_atomatom_2016}, or other dielectric structures, like atomic lattices with sub-wavelength spacing~\cite{bettles_enhanced_2016, shahmoon_cooperative_2017, asenjo-garcia_exponential_2017, ruiSubradiantOpticalMirror2020,srakaewSubwavelengthAtomicArray2023a, reitz_cooperative_2022, ruostekoski_cooperative_2023}. These platforms offer a wide range of applications, including the generation of super- and subradiant states~\cite{masson_many-body_2020, ferioli_storage_2021, rubies-bigorda_photon_2022, holzinger_control_2022, rubies-bigorda_dynamic_2023, tiranovCollectiveSuperSubradiant2023}; non-classical states of light~\cite{williamson_superatom_2020, cidrim_photon_2020, parmee_signatures_2020, ferioli_non-gaussian_2024}; quantum simulation and information processing~\cite{shah_quantum_2024}; and sensing~\cite{rubies-bigorda_dynamic_2023, ye_essay_2024}.

In this Letter, we propose a versatile measurement protocol to exploit two quantum emitters, dissipatively coupled through their embedding in a 2D atomic array~\cite{masson_atomic-waveguide_2020, patti_controlling_2021, buckley-bonanno_optimized_2022}, for sensing applications. We introduce the term \emph{array QED} to denote this conceptual framework, drawing parallels to concepts from cavity~\cite{walther_cavity_2006}- and waveguide~\cite{sheremet_waveguide_2023} QED. The utilization of quantum platforms for quantum sensing capitalizes on a core feature of quantum systems: their susceptibility to external disturbances. While these effects are detrimental to quantum cryptography and quantum computing, they present an opportunity for highly sensitive measurements of electric and magnetic fields~\cite{bianNanoscaleElectricfieldImaging2021}, time and frequency~\cite{bloomOpticalLatticeClock2014}, rotations~\cite{campbellRotationSensingTrapped2017}, temperature, and pressure~\cite{degenQuantumSensing2017, hsieh_imaging_2019}. 

In our approach, we measure the population transfer between two strongly-coupled distant emitters, exploiting their separation to perform local excitation at one emitter and local population readout at the other. In analogy to waveguide QED and cavity QED, array QED can benefit from collective enhancement, leading to much stronger atom-photon interactions compared to free space. We show that this enhancement enables the measurement of minute frequency shifts.

Through an analytic description of the population transfer dynamics, we show how this cooperative enhancement can be used for sensing, while remaining robust to certain types of noise.

\emph{Model.}---A pair of cooperatively coupled quantum emitters, labelled $1$ and $2$, are embedded in a 2D sub-wavelength atomic array as shown in Fig.~\ref{fig:System}a. These emitters have decay rates $\gamma_{1,2}$ and resonance frequencies $\omega_{1,2} = 2\pi c/\lambda_0^{1,2}$, where $\lambda_0^{1,2}$ is the transition wavelength of emitters $1$ and $2$.
The lattice is a large two-dimensional atomic array with a sub-wavelength lattice spacing $a < \lambda_0^{1,2}$. In this setup, the atoms interact through light-induced dipole-dipole interactions.

After adiabatically eliminating the lattice, the system can be modeled by the effective non-Hermitian Hamiltonian:
\begin{align}
\label{eq:heff}
\mathcal{H}_{\mathrm{eff}} &=  \left(
\begin{array}{cc}
 \Omega_1 - i \Gamma_1/2 & \sqrt{\gamma_1\gamma_2} \kappa \\
\sqrt{\gamma_1\gamma_2} \kappa & \Omega_{q} - i \Gamma_2/2 \\
\end{array}
\right),
\end{align}
which is an effective system comprised of only the two impurities (see~\fref{fig:System}b) realized by collective coupling induced by the lattice. 
The coherent and dissipative interaction strengths between the atoms are determined via the real and imaginary part of the Green's tensor for a point dipole in free space (see Supplementary Information~\cite{supplement}), physically corresponding to a modification of the effective frequencies
$\Omega_1 \equiv \gamma_1 \re\{\Sigma\}, \Omega_2 \equiv \Delta +\gamma_2 \re\{\Sigma\}$,
and effective decay rates 
$\Gamma_1 \equiv \gamma_1 \left(1- 2 \im\{\Sigma\} \right) = \gamma_1 \Gamma_\mathrm{coop}$ and $\Gamma_2 \equiv \gamma_2 \left(1- 2 \im\{\Sigma\} \right) = \gamma_2 \Gamma_\mathrm{coop}$,  where $\Delta= \omega_1-\omega_2$ is the detuning between the two impurities.
The parameter
\begin{align}
\Gamma_\mathrm{coop} \equiv 1-2 \im(\Sigma),    
\end{align}
is the cooperativity factor in array QED that describes the lengthening of the excited state lifetimes for the impurities ($0<\Gamma_\mathrm{coop}<1)$, and is thus a key parameter in our protocol, allowing the system to retain coherence for longer. 
The self-energy $\Sigma$ describes the modification to the atom's energy as a result of its own field and the fields of the surrounding atoms, while the coupling strength $\kappa$ describes the strength of interaction between neighboring atoms. Note that energy shifts resulting in $\Sigma$ usually diminish the sensitivity of lattice based sensors~\cite{kramer_optimized_2016}. The protocol presented below is robust against these cooperatively induced shifts.
In general, the system dynamics of a dissipative quantum system should be modelled in the master equation formalism. However, in the single-excitation subspace, and without external driving, the quantum jump terms in the master equation can be neglected~\cite{olmosOutofequilibriumEvolutionKinetically2014} and the system dynamics are fully described by the effective Hamiltonian \eqref{eq:heff}.

\begin{figure}[tp!]
\includegraphics[width=0.8\linewidth, trim=0cm 0.0cm 0.0cm 0.0cm,clip]{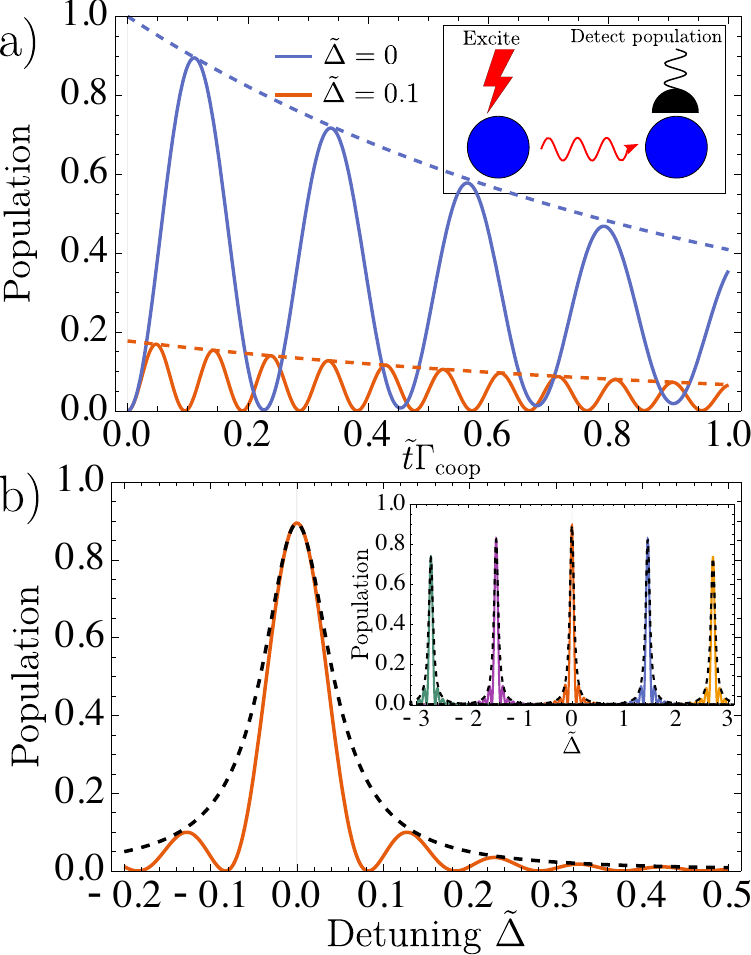}
	\caption{
	 a) The population of the initally un-excited impurity $1$ as a function of the time $\tilde t \Gamma_\mathrm{coop}$, where $\tilde t= \sqrt{\gamma_1 \gamma_2} t$ [see Eq.~\eqref{eq:population}] for a $10 \times 10$ lattice, with lattice spacing $a=0.3\lambda_0^s$ and impurity distance $d=2a$. The blue curve corresponds to maximal population transfer at $\tilde{\Delta} = 0$. A small change in detuning ($\tilde{\Delta}=0.1$) diminishes the achievable population transfer (red curve).
 The dashed lines show the overall decay of the population (proportional to $\Gamma_\mathrm{coop}$) without the transition between $1$ and $2$.
    b) The maximum population as a function of detuning for the same parameters as a) above. 
    In the inset, the population is shown for values of $R=\gamma_1/\gamma_2$ from left to right $R= (1/30,1/10,1,10,30)$.
    The height and width of the quasi-Lorenzian $|c_{1_\mathrm{max}}|^2(\Delta) \approx  \gamma_1 \gamma_2 \left|\frac{\kappa}{S}\right|^2 e^{-\frac{\pi \bar \gamma \Gamma_\mathrm{coop}}{2\re(S)}}$ (dashed lines) lowers and widens as we move away from the ratio $R=\gamma_1/\gamma_2 = 1$. 
}
	\label{fig:impurity_dynamics}
\end{figure}

It may appear that this Hamiltonian produces standard damped Rabi oscillations (see also \fref{fig:impurity_dynamics}a). However, by taking into account both dissipative and cooperative coupling, the effective Hamiltonian acquires additional imaginary components, extending the physics beyond the conventional damped Rabi regime.


\emph{Sensing protocol}.---The key idea of the proposed protocol is illustrated in Fig.~\ref{fig:impurity_dynamics}a. After an initial excitation of impurity $2$, the population in impurity $1$ is detected after some time $t_0$. 
The population in the detector atom $1$ is highly sensitive to small changes in the relative resonance frequencies between the quantum emitters, described by the detuning $\tilde \Delta = \Delta/\sqrt{\gamma_1 \gamma_2}$ (see~\fref{fig:impurity_dynamics}a,b.).
Throughout this Letter, we will use the tilde ($\sim$) to denote a renormalization of a parameter by the factor $\sqrt{\gamma_1 \gamma_2}$, where $\gamma_{i}$ has units of frequency.
As a result of this sensitivity, an external disturbance to impurity $2$ that results in a shift of the atomic resonance frequency, e.g., a magnetic or gravitational field, can be detected in the dynamics of impurity $1$.
This is the key feature of our proposed sensing protocol, since it allows the measurement of frequency shifts via population measurements in impurity $1$. 

The system dynamics are governed by the differential equations $(\dot c_1, \dot c_2)^T= \mathcal{H}_\mathrm{eff} \cdot (s,q)^T$, the solutions to which are superpositions of exponentials of the eigenvalues of $H_\mathrm{eff}$, which are 
$\omega_{\pm} =  \bar \Omega  -i \frac{\bar \gamma \Gamma_\mathrm{coop}}{2} \pm S$
with the quasi-Rabi frequency
\begin{align}
\label{eq: quasi-rabi}
    S(\Delta)=\sqrt{ \frac{1}{4} \left(\Delta_0-\Delta - i \frac{\Gamma_\mathrm{coop}}{2} (\gamma_1-\gamma_2)\right)^2 +  \gamma_1 \gamma_2\kappa^2}.
\end{align}
The resonant detuning $\Delta_{0} = \re(\Sigma ) \left(\gamma_1-\gamma_2\right)$,
corresponds physically to when the two impurity frequencies are equal $\Omega_1 = \Omega_2$.
In the inset of Figure \ref{fig:impurity_dynamics}b, we can see that the resonance peaks shift as the ratio $R=\gamma_1/\gamma_2$ is changed. The location of the maximum of these peaks is given by $\Delta_0$.
However, since operation with $\gamma_1 =\gamma_2$ is globally optimal (see Supplementary Information~\ref{SM: optimalR}), both the resonant detuning $\Delta_0$ and the term containing $\Gamma_\mathrm{coop}$ drop out of the quasi-Rabi frequency~\eqref{eq: quasi-rabi}.

As stated before, after one impurity is excited, the population of the other impurity is strongly dependent on the detuning $\Delta$ between their frequencies. 
Assuming impurity $2$ is initially excited, the population of impurity $1$ at time $t$ is given as
\begin{align}
\label{eq:population}
|c_1|^2 &= \frac{\gamma_1 \gamma_2| \kappa|^2}{2|S|^2} e^{-t \bar{\gamma }\Gamma_\mathrm{coop}} \left[\cosh(2 t \im(S))-\cos(2 t \re(S))\right],
\end{align}
where $S = S(\Delta)$ shows where this dependence on detuning enters the dynamics. $\bar \gamma = (\gamma_1 + \gamma_2)/2$ is the mean decay rate. 
Since the cosh term diverges for $t\to \infty$, we require $\bar \gamma \Gamma_\mathrm{coop} > 2 \im(S)$ (see Supplementary Information \ref{SM:constraints}).
As we will later discuss, the cosh term and the decay term containing $\Gamma_\textrm{coop}$ are the key terms that lengthen the system's coherence time and allows for enhanced sensitivity.


The dynamics of $c_1(t)$ differ from the well known behavior of laser driven atoms performing Rabi oscillations. In our system, $S(\Delta)$ takes on the analogous role of the Rabi frequency, with the key difference being that both the quasi-detuning $\Delta_\mathrm{quasi} = \Delta -\Delta_0 - i \frac{\Gamma_\mathrm{coop}}{2}(\gamma_2-\gamma_1)$ given by the difference between the diagonal elements of $\mathcal{H}_\mathrm{eff}$, and consequently the quasi-Rabi frequency $S(\Delta)$ have imaginary components. These imaginary components are a consequence of the decay channels available due to the cooperative effects induced via the surrounding lattice atoms.

Due to the non-Hermitian effective Hamiltonian~\eqref{eq:heff} the eigenvalue spectrum of $H$ admits exceptional points~\cite{heiss_exceptional_2004, wiersigEnhancingSensitivityFrequency2014, ozdemir_paritytime_2019}, which have been previously investigated as candidates for quantum sensing protocols~\cite{wiersig_prospects_2020, wiersig_review_2020}. For the protocol presented below, the exceptional points are not optimal points for sensing, so are not the focus of this work (see Supplementary Information~\ref{SM:exceptional_points}).   


We will now precisely define how we quantify the sensitivity of the protocol. 
Since the protocol requires the detection of single photons at the location of impurity $1$, each measurement of the population in impurity $1$ is a Bernoulli trial (the detector clicks or it doesn't) where the probability of success is binomially distributed with probability $p = |c_1(t_0)|^2$, i.e., we take samples from the response curve in Figure~\ref{fig:impurity_dynamics}b. From these measurements, one can estimate the quantity $|c_\mathrm{signal}|^2$, which is the change in the population due to some additional detuning $\Delta_\mathrm{signal}$. Since the population changes with $\Delta_\mathrm{signal}$, by measuring $|c_1|^2$, we can obtain an estimate of $\Delta_\mathrm{signal}$ from $|c_\mathrm{signal}|^{2}$.

Since $\Delta_\mathrm{signal}$ inherits the binomial distribution from $|c_\mathrm{signal}|^2$, we can propagate the measurement uncertainty to obtain an estimate (see Supplementary Information~\ref{SM:error_propagation} for more details)
\begin{align}
\label{eq:sigma_signal}
    \sigma_{\Delta_\mathrm{signal}} = \frac{\sigma_{|c_1|^2}}{\left|\frac{\partial |c_1|^2}{\partial \Delta}\right|},
\end{align}
where the derivative is evaluated at an added detuning $\Delta_\mathrm{add}$, which is an additional degree of freedom that allows for further optimization of the system.
Since external disturbances will alter the optimal operating point, the additional control afforded by $\Delta_\mathrm{add}$ gives the protocol additional flexibility to operate at the new optimum. 
The measurements of the population $|c_1|^2$ are distributed binomially and so have the well-known variance $\sigma^2 = p(1-p)$, where $p=|c_1|^2$.
A large derivative $\partial |c_1|^2/\partial \Delta$ means a small change in $\Delta_\mathrm{signal}$ leads to a large change in the $|c_1|^2$, reducing the uncertainty in $\Delta_\mathrm{signal}$ for a given uncertainty in $p$. 
The derivative encodes the intuitive physical understanding that the `narrowness' of \fref{fig:impurity_dynamics}b leads to a more sensitive system. 
Additionally, the population variance $\sigma_{|c_1|^2}$ is  minimized at $p=1$, so given two systems where the derivative is equal, it is advantageous to choose system parameters that lead to the largest population transfer. 
However, any change in the population itself is likely to also affect the behavior with respect to $\Delta$, and thus to fully optimize the sensing, we need to consider the interplay between $p$ and this derivative. 

To summarize, the observable of interest is the signal detuning $\Delta_\mathrm{signal}$.
For a certain frequency shift, which is generated via an external \emph{local} perturbation (e.g., a magnetic field at the location of impurity $2$). Since the system is highly sensitive to changes in $\Delta$, the protocol is able to detect small external disturbances $\Delta_\mathrm{signal}$. We quantify this sensitivity by computing the uncertainty $\sigma_{\Delta_\mathrm{signal}}$  [see \ref{fig:impurity_dynamics}b and \eqref{eq:sigma_signal}]. We will henceforth drop the subscript $\Delta_\mathrm{signal}$ and refer to the uncertainty simply as $\sigma$, the standard deviation of a measurement of $\Delta_\mathrm{signal}$.
A question now remains: how do our system parameters, the lattice spacing $a$, the added detuning $\Delta_\mathrm{add}$ and the measurement time $t_0$ affect this measurement sensitivity?

 \begin{figure}[tp!]
\includegraphics[width=0.9\linewidth, trim=0.05cm 0.2cm 0.05cm 0.05cm,clip]{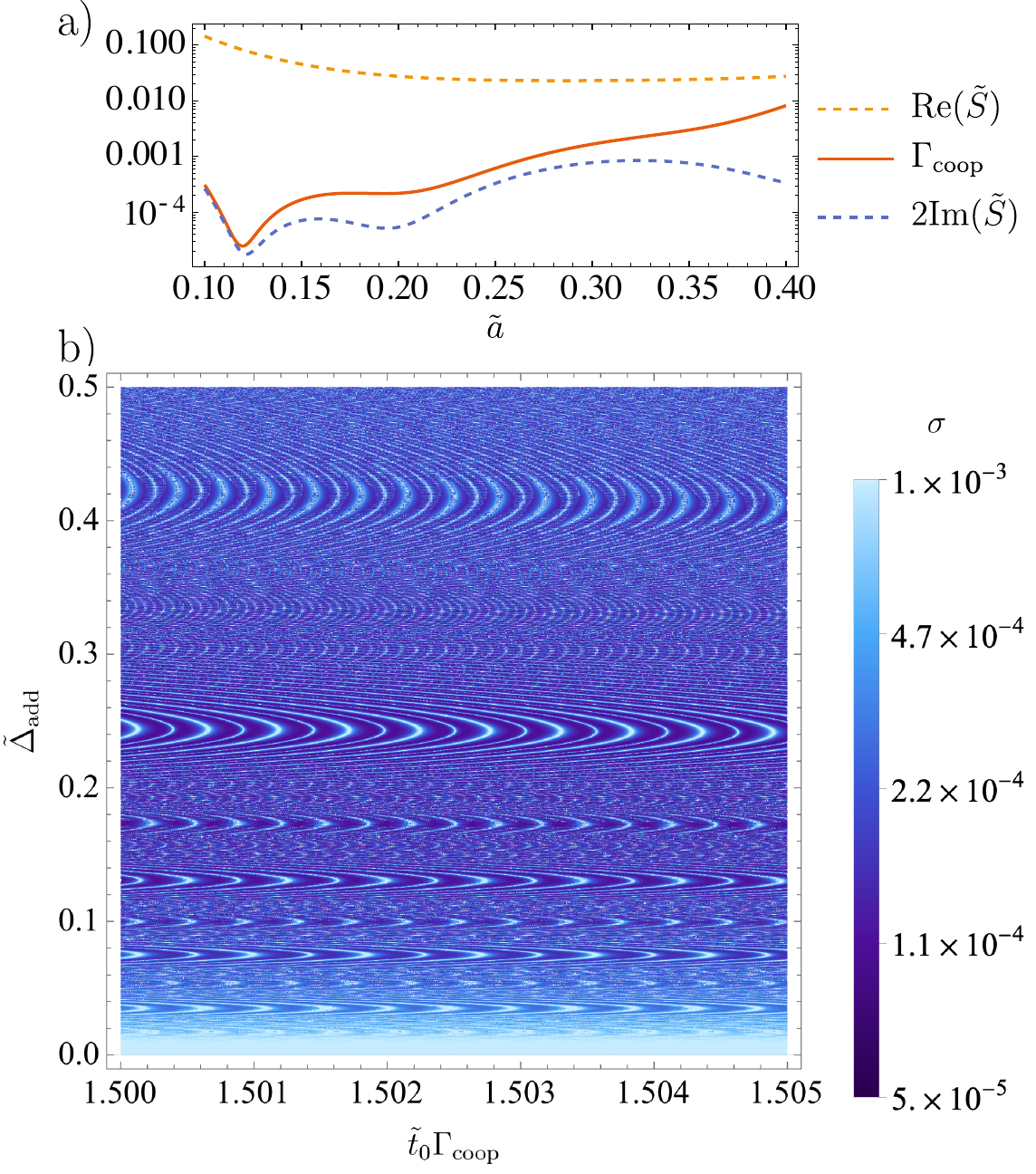}
	\caption{ 
	(a) The cooperative enhancement factor $\Gamma_\mathrm{coop}$, as well as the real and twice the imaginary part of the quasi-Rabi frequency as a function of the lattice spacing $\tilde a$. 
 (b)
 Standard deviation of measurements of $\Delta_\mathrm{signal}$ as a function of the measurement time $\tilde t_0 \Gamma_\mathrm{coop}$, and $\tilde \Delta_\mathrm{add}$ where $\tilde t_0 = \sqrt{\gamma_1 \gamma_2} t_0$ and $\tilde \Delta_\mathrm{add} = \Delta_\mathrm{add}/\sqrt{\gamma_1 \gamma_2}$.
 The resonances are a result of the derivative reaching an inflection point, leading to a divergence in Eq.~\eqref{eq:sigma_signal}.
 We set $\gamma_1 = \gamma_2$ for this plot as it represents the universally optimal value for this protocol (see Supplementary Information~\ref{SM: optimalR}).
}	\label{fig:sigma_vs}
\end{figure}

The lattice spacing changes the timescale of the dynamics, where a smaller lattice spacing means faster population exchange between the impurities. This makes physical sense: the smaller the lattice spacing, the smaller the physical distance the excitation needs to traverse, and assuming equal decay rates and propagation velocity, the population will oscillate more rapidly. 
This is reflected in the real part of the quasi-Rabi frequency, which we show in \fref{fig:sigma_vs}a. As $a$ deceases, $\re(\tilde S)$ increases monotonically, leading to a corresponding increase in the oscillation frequency, in analogy with Rabi oscillations. 
An enhanced oscillation rate results in a larger change in the population given some change to the detuning, $\Delta$, thus improving the sensitivity. 

There are two decay terms in \eqref{eq:population}: one cosh term containing $\im(S)$ and one containing $\Gamma_\mathrm{coop}$. 
To have the best sensitivity, the system should retain coherence as long as possible, i.e., we would like the overall decay to be as small as possible.
Since the cosh term contributes to an overall growth of the population for $\im(S)>0$, it is optimal to satisfy the condition $\Gamma_\mathrm{coop} > 2 \im(\tilde S)$ as close to equality as possible.
However, the overall decay is dominated by $\Gamma_\mathrm{coop}$, which should also be minimized. 
The optimum for both conditions lies at $\tilde a \sim 0.12$ (see \fref{fig:sigma_vs}a), which is a global optimum for the sensitivity with respect to the lattice spacing. 
Generally, decreasing the lattice spacing has the dual effect of increasing the oscillation frequency of the dynamics themselves, whilst decreasing the decay, so overall the system oscillates more rapidly and does so for longer, both of which improve the sensitivity of the protocol. 

In \fref{fig:sigma_vs}b, $\sigma$ displays alternating broad and narrow resonances as a function of the measurement time $t_0$.
The broad resonances in this plot stems from the population hitting a maximum, leading to a zero derivative and a subsequent divergence in $\sigma$. 
The narrower resonances correspond to the population reaching a minimum.
For a given $t_0$, the added detuning $\Delta_\mathrm{add}$ serves to move these resonances around, resulting in the repeating structures that can be seen in \fref{fig:sigma_vs}b.
An added detuning $\Delta_\mathrm{add}$ increases the effective oscillation frequency, and thus will tend to increase the derivative $\partial(|c_1|^2)/\partial \Delta$. However, a detuning leads to decreased coupling efficiency, reducing the population and increasing the variance. 
The optimal choice for $\Delta_\mathrm{add}$ lies in the dark horizontal bands \fref{fig:sigma_vs}b, avoiding the resonances in $t_0$.
Physically, this corresponds to choosing $\Delta_\mathrm{add}$ and $t_0$ such that the population is changing rapidly with $\Delta$, maximizing the derivative $\partial |c_1|^2/\partial \Delta$.
The possible choices of $t_0$ are expanded greatly due to the decay suppressing effects of $\Gamma_\mathrm{coop}$ and $\im(S)$, as the population can remain high even after several cycles of oscillations. Finally, we note that while the discussion above focused on initial excitation of a single impurity, the protocol can also be applied to alternative initializations by tracking the population exchange between the two impurities over time.

\begin{figure}[tp!]
\includegraphics[width=\linewidth, trim=0.00cm 0.00cm 0.01cm 0.0cm,clip]{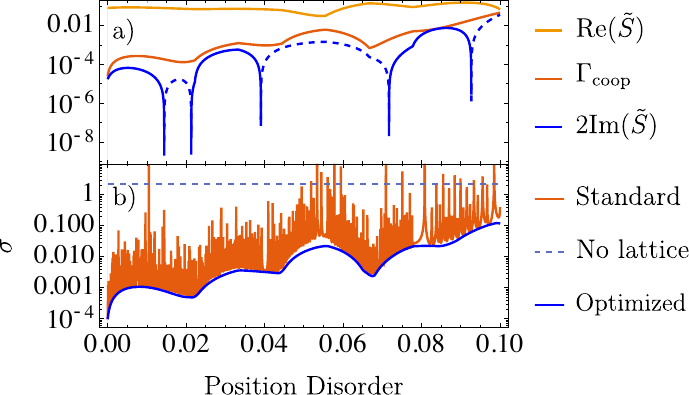}
	\caption{ 
 a) The modified $\Gamma_\mathrm{coop} = 1-\im(\Sigma_1)-\im(\Sigma_2)$ as well as the real and imaginary parts of the quasi-Rabi frequency $\tilde S = S/\sqrt{\gamma_1 \gamma_2}$. The dashed blue lines represent the negative values of $2 \im(\tilde S)$.
 b) The measurement standard deviation $\sigma$ plotted for a lattice initialized with amount of Gaussian disorder. We show $\sigma$ when the optimal parameters at zero disorder are unchanged (Standard), as well as the re-optimized curve, which involves taking into account the new lattice to find a new set of optimal parameters at each value of the disorder. 
\label{fig:sigma_disorder}}	
\end{figure}

\emph{Robustness to noise.}---
In realistic experiments, the lattice will have  imperfections due to both atomic motion and position. 
We model these imperfections by randomly sampling the positions of the lattice atoms from a Gaussian distribution centered about the old position but with a variance that we call ``position disorder" (this is to prevent confusion with the $\sigma_{\Delta_\mathrm{signal}}$ already in use).  
This procedure breaks the lattice symmetry and requires that we include different self-energies $\Sigma_1$ and $\Sigma_2$, for each impurity. 
Now $\Gamma_\mathrm{coop} = 1-\im(\Sigma_1)-\im(\Sigma_2)$ and $\Delta_0 = \gamma_1\re(\Sigma_1)-\gamma_2 \re(\Sigma_2)$.
In \fref{fig:sigma_disorder}a), we see that as position disorder grows, $\Gamma_\mathrm{coop}$ grows along with it, and as is reflected in \fref{fig:sigma_disorder}b), so does the sensitivity of the protocol, which follows the variations in $\Gamma_\mathrm{coop}$ closely. 
As before, it is the combination of a small $\Gamma_\mathrm{coop}$ that is close to (but larger than) a small value of $2\im(\tilde S)$ that will lead to optimal results for the sensing protocol. 
The real part of the quasi-Rabi frequency stays roughly constant throughout the range of position disorders, indicating that the oscillation rate of the population transfer is not strongly affected by position disorder. 
In \fref{fig:sigma_disorder}b), we show the effect of the position disorder on $\sigma$.
Changes in the lattice can be detected by the measurements of the population, giving the flexibility to find a new optimal operating point. 
Since $\Gamma_\mathrm{coop}$ stays relatively small even for large position disorder, the performance of the re-optimized protocol significantly outperforms the case of two impurities without a lattice (shown by the dashed line), for all values of the position disorder. This illustrates that the protocol presented above is robust against lattice disorder and cooperative enhancement prevails in this case.


\emph{Conclusions and Outlook.}---
We introduced a quantum sensing protocol based on coherent photon exchange between distant quantum emitters. 
We showed that local population measurements can be used to detect frequency shifts due to the strong sensitivity of the photon transfer efficiency on the relative detuning between the emitters. 
The population transfer is strongly enhanced by the collective effects present in the system. This is in contrast to other lattice based metrology applications, which are in general harmed by such effects via dephasing and cooperative shifts.
While the discussion in this Letter was based on an implementation with impurities embedded in a cooperative array, the protocol is, in principle, generalizable to related platforms (waveguide coupled quantum dots, photonic bandgap materials, solid state based implementations, NV centers etc).

The introduced approach opens up exciting avenues for cooperatively enhanced quantum sensing. The major challenge to overcome is local single photon readout (which is becoming feasible in, e.g., modern optical tweezer based setups).

Additionally, the population dynamics of the two impurities could be used to monitor the lattice dynamics, forming the basis for a new tool for device characterization. To further improve the sensitivity, another potential future direction is to extend the study to systems beyond the single excitation regime, where strong photon non-linearities will occur. 

\begin{acknowledgments}
We would like to thank E. Kalberer for assistance in the production of figures.
O.S. acknowledges funding from the European Union’s Horizon 2020 research and innovation program under the Marie Skłodowska-Curie grant agreement No. 801199. SFY acknowledges funding from the NSF via PHY-2207972, the CUA PFC PHY-2317134, and the Q-SEnSE QLCI OMA-2016244. We gratefully acknowledge discussions with Anders S. S\o rensen. 
\end{acknowledgments}


\bibliographystyle{apsrev4-2}
\bibliography{cooperative_sensing_refs.bib}

\pagebreak

\clearpage
\onecolumngrid
\begin{center}

\newcommand{\beginsupplement}{%
        \setcounter{table}{0}
        \renewcommand{\thetable}{S\arabic{table}}%
        \setcounter{figure}{0}
        \renewcommand{\thefigure}{S\arabic{figure}}%
     }
\textbf{\large Supplemental Material}
\label{SupplementaryMaterial}
\end{center}
\newcommand{\beginsupplement}{%
        \setcounter{table}{0}
        \renewcommand{\thetable}{S\arabic{table}}%
        \setcounter{figure}{0}
        \renewcommand{\thefigure}{S\arabic{figure}}%
     }

\setcounter{equation}{0}
\setcounter{figure}{0}
\setcounter{table}{0}   
\setcounter{page}{1}
\makeatletter
\renewcommand{\theequation}{S\arabic{equation}}
\renewcommand{\thefigure}{S\arabic{figure}}
\renewcommand{\bibnumfmt}[1]{[S#1]}
\renewcommand{\citenumfont}[1]{S#1}
\vspace{0.8 in}
\newcommand{\D}{\Delta}
\newcommand{\tD}{\tilde{\Delta}}
\newcommand{\K}{K_{PP}}
\newcommand{\bn}{\bar{n}_P}
\newcommand{\G}{\Gamma}
\newcommand{\LH}{\underset{L}{H}}
\newcommand{\HL}{\underset{H}{L}}
\vspace{-1in}

\section{Realisation of effective Hamiltonian}
Below we show how the effective Hamiltonian \eqref{eq:heff} can be obtained via adiabatic elimination of the lattice Hamiltonian in the appropriate rotating wave approximation. 
\label{SM: Realisation of effective Hamiltonian}
We model an array of $\ell$ atoms arranged in a lattice, as well as two impurities with annihilation operators $a_1,a_2$. 
We have a space of size $N$, where the first $\ell$ entries correspond to the lattice atoms, and the next two entries correspond to the impurities, i.e., $\ell = N-2$. 

In real space, the Hamiltonian is split into
\begin{align}
H_{latt} &= \sum_m^{\ell} \left( \omega_L - i \frac{\gamma_L}{2}\right) \sigma_m^{\dag}\sigma_m + \gamma_{L}\sum_{m \neq n}^{\ell} \left( J_{mn} - i \frac{\Gamma_{mn}}{2} \right) \sigma^{\dag}_m \sigma_n
\label{eqn:H_latt}
\end{align}
We have also the interaction between the lattice and the impurities (or defect)
\begin{align}
H_{latt, d} = \sqrt{\gamma_L \gamma_d} \sum_m^{\ell} \left[\left( J_{d m} - i \frac{\Gamma_{dm}}{2}\right) a_d^{\dag} \sigma_{m}+ \left( J_{m d} - i \frac{\Gamma_{m d}}{2}\right) \sigma_m^{\dag} a_d  \right]
\label{eqn:H_latt_d}
\end{align}
where $d = 1,2$ indicate the impurity atoms and $\sigma_z^{(d)}$ indicates a $\sigma_z$ matrix acting on site $d$.  
We have also the impurity interaction that has the same form as the lattice atoms 
\begin{align}
H_{21} = \sqrt{\gamma_2 \gamma_1} \left[\left( J_{21} - i \frac{\Gamma_{21}}{2}\right) a_2^{\dag} a_1 + \left( J_{12} - i \frac{\Gamma_{12}}{2}\right) a_1^{\dag} a_2 \right]
\label{eqn:H_qs}
\end{align}
The bare impurity dynamics is governed by
\begin{align}
H_{d d} = \sum_{d} \left( \omega_{d} -i \frac{\gamma_d}{2} \right) a_d^{\dag} a_d
\label{eqn:H_d_d}
\end{align}
Due to resonant dipole-dipole interactions, we obtain collective coupling $J_{ij}$ and decay $\Gamma_{ij}$ rates that are defined as 
\begin{align}
\label{eq: collectiveDissipativeCoupling}
\sqrt{\gamma_i\gamma_j}J_{ij}(\mathbf{r}_i,\mathbf{r}_j) &= -\frac{3 \pi \sqrt{\gamma_i \gamma_j}}{\omega} \mathbf{d}^{\dag}_i\cdot \re\left[ \mathbf{G}(\mathbf{r}_{ij},\omega)\right] \cdot \mathbf{d}_j
\\ \sqrt{\gamma_i \gamma_j}\Gamma_{ij}(\mathbf{r}_i,\mathbf{r}_j) &= -\frac{6 \pi \sqrt{\gamma_i \gamma_j}}{\omega} \mathbf{d}^{\dag}_i\cdot \im\left[ \mathbf{G}(\mathbf{r}_{ij},\omega)\right] \cdot \mathbf{d}_j
\end{align}
for the free space Green's tensor $\mathbf{G}(\mathbf{r}_{ij},\omega)$ with components
\begin{equation} 
    G_{\alpha\beta} (\mathbf{r}_{ij}, \omega) = \frac{e^{i\omega r}}{4\pi r} \left[ \left( 1 + \frac{i}{\omega r} - \frac{1}{\omega^2 r^2} \right) \delta_{\alpha\beta} 
    - \left( 1 + \frac{3i}{\omega r} - \frac{3}{\omega^2 r^2} \right) \frac{r_{ij,\alpha} r_{ij,\beta}}{r^2} \right] - \frac{\delta(\mathbf{r}_{ij})}{3\omega^2} \delta_{\alpha\beta},
    \label{eqn:Green}
\end{equation}
an atomic separation $\mathbf{r}_{ij} = \mathbf{r}_i- \mathbf{r}_j$, $r\equiv |\mathbf{r}_{ij}|$, individual atomic decay rates $\gamma_i, \gamma_j$ and atomic dipole moments $\mathbf{d}_{i,j}$. 
\subsection{Momentum space representation}
Writing $\sigma_m = \frac{1}{\sqrt{\ell}} \sum_{\mathbf{k}=1}^{\ell} e^{\frac{2\pi}{\ell} i \mathbf{k} \cdot \mathbf{r}_m}$, moving to the rotating frame with frequency $\omega_1$ and requiring that we are in the single-excitation subspace, we get that
\begin{align}
H_{latt} &=  \gamma_L\sum_{\mathbf{k}} \left[\frac{\delta_{1L}}{\gamma_L}- \frac{i}{2} +  J_L(\mathbf{k}) - \frac{i}{2}\Gamma_L(\mathbf{k}) \right] \sigma_{\mathbf{k}}^{\dag} \sigma_{\mathbf{k}}
\end{align}

\begin{align}
H_{latt,d} = \sqrt{\gamma_L\gamma_d} \sum_{\mathbf{k}} \left[ \left( J_d(\mathbf{k}) - \frac{i}{2}\Gamma_d(\mathbf{k}) \right) a_d^{\dag}\sigma_{\mathbf{k}}  + \left( J^{*}_d(\mathbf{k}) - \frac{i}{2}\Gamma^{*}_d(\mathbf{k}) \right) \sigma_{\mathbf{k}}^{\dag}d  \right]
\end{align}
\begin{align}
H_{21} = \sqrt{\gamma_1 \gamma_2} \left( J_{21} - i \frac{\Gamma_{21}}{2}\right) a_2^{\dag} a_1 + \left( J_{12} - i \frac{\Gamma_{12}}{2}\right) a_1^{\dag} a_2 
\end{align}

\begin{align}
H_{11} &= -i \frac{\gamma_1}{2} a_1^{\dag} a_1
\\
H_{22} &= \left( \Delta -i \frac{\gamma_2}{2} \right) a_2^{\dag}a_2
\end{align}

Where $\Delta = \omega_2- \omega_1$, $\delta = \omega_L - \omega_1$ and where $\sigma_{\mathbf{k}} \to e^{-i\omega_1 t} \sigma_{\mathbf{k}}$, $a_1 \to e^{-i\omega_1 t}a_1$ and $a_2 \to e^{-i\omega_1 t}a_2$.

After adiabatic elimination of the lattice, we obtain 
\begin{align}
   \dot c_1 &=  -i\left(\gamma_1\Sigma_1- i \frac{\gamma_1}{2}\right) c_1 -i\sqrt{\gamma_1\gamma_2}\kappa_1 c_2
\\ \dot c_2 &=  -i\left(\Delta + \gamma_2\Sigma_2- i \frac{\gamma_2}{2}\right) c_2 -i \sqrt{\gamma_1\gamma_2}\kappa_2 c_1
\end{align}

Where we define the self-energies
\begin{align}
\Sigma_{1}&\equiv\sum_{\mathbf{k}} \frac{(J_1(\mathbf{k})-\frac{i}{2}\Gamma_1(\mathbf{k}))(J^{*}_1(\mathbf{k})-\frac{i}{2}\Gamma^{*}_1(\mathbf{k}))}{\delta/\gamma_L+J_L(\mathbf{k}) - \frac{i}{2}\Gamma_L(\mathbf{k}) - i/2}\\
\Sigma_2&\equiv\sum_{\mathbf{k}} \frac{(J_2(\mathbf{k})-\frac{i}{2}\Gamma_2(\mathbf{k}))(J^{*}_2(\mathbf{k})-\frac{i}{2}\Gamma^{*}_2(\mathbf{k}))}{\delta/\gamma_L+J_L(\mathbf{k}) - \frac{i}{2}\Gamma_L(\mathbf{k}) - i/2}
\label{eqn:self-energy_k}
\end{align}
and the effective coupling strengths 
\begin{align}
\kappa_1 & \equiv \left(  J_{12} - i \frac{\Gamma_{12}}{2} + \sum_{\mathbf{k}} \frac{(J_1(\mathbf{k})-\frac{i}{2}\Gamma_1(\mathbf{k}))(J^{*}_2(\mathbf{k})-\frac{i}{2}\Gamma^{*}_2(\mathbf{k}))}{\delta/\gamma_L+J_L(\mathbf{k}) - \frac{i}{2}\Gamma_L(\mathbf{k}) - i/2}\right)
\\ \kappa_2 &\equiv\left(  J_{21} - i \frac{\Gamma_{21}}{2} + \sum_{\mathbf{k}} \frac{(J_2(\mathbf{k})-\frac{i}{2}\Gamma_2(\mathbf{k}))(J^{*}_1(\mathbf{k})-\frac{i}{2}\Gamma^{*}_1(\mathbf{k}))}{\delta/\gamma_L+J_L(\mathbf{k}) - \frac{i}{2}\Gamma_L(\mathbf{k}) - i/2}\right)
\label{eqn:couplings_k}
\end{align}

Now calling  $\ket{\psi} = \begin{pmatrix} c_1\\c_2 \end{pmatrix}$ and noting that $\kappa_1 = \kappa_2 \equiv \kappa$ and $\Sigma_1= \Sigma_2 \equiv \Sigma$, these equations can be written in terms of an effective (non-Hermitian) Hamiltonian

\begin{align}
\mathcal{H}_{\mathrm{eff}} &= \left(
\begin{array}{cc}
 -\frac{i \gamma_1}{2}+\gamma_1\Sigma & \sqrt{\gamma_1\gamma_2}\kappa \\
\sqrt{\gamma_1\gamma_2} \kappa & -\frac{i \gamma_2}{2}+\Delta+ \gamma_2\Sigma \\
\end{array}
\right) \equiv \left(
\begin{array}{cc}
 \Omega_1 - i \Gamma_1/2 & \sqrt{\gamma_1\gamma_2} \kappa \\
\sqrt{\gamma_1\gamma_2} \kappa & \Omega_{q} - i \Gamma_2/2 \\
\end{array}
\right),
\label{eqn:Heff}
\end{align}
which is \eqref{eq:heff} in the main text. 
\subsection{Constraints on $\Sigma$ and $\kappa$}
\label{SM:constraints}
The above derivation requires the computation of the self-energy $\Sigma$ and the coupling rate $\kappa$ if the dynamics of the effective Hamiltonian~\eqref{eq:heff} are to be realised. 
Below we discuss the constraints and physical meaning of these parameters. 

The real part of $\Sigma$ is responsible for setting an energy scale, and so does nothing whatsoever to the overall dynamics, as the only relevant quantity affecting dynamics is the energy difference. To achieve the longest lifetime, the requirement is $\im\{\Sigma\} = 1/2$, such that the decay terms $\Gamma_1$ and $\Gamma_2$ are zero and thus the decay time diverges. 
However due to the cosh term, at $\im\{\Sigma\}=1/2$ we have a positive overall decay, such that  $\lim_{t\to \infty} |c_1(t)|^2 \to \infty$, i.e., the population diverges and is thus unphysical.
In general, for us to have physical parameters, we need to satisfy
\begin{align}
\im\{\Sigma\} < \frac{1}{2} \, \, \, \mathrm{and} \, \, \, \bar{\gamma} \Gamma_\mathrm{coop} > 2\im\{S \}, 
\end{align}
where the second inequality comes from ensuring that $\lim_{t \to \infty} |c_1(t)|^2< 1$.
Satisfying the second inequality puts restrictions on $\kappa$. 
Assuming $\gamma_1= \gamma_2 \equiv \gamma$, and $\Delta =0$  we have $S =\sqrt{\gamma ^2 \kappa ^2}$, and
\begin{align}
\im(S)=  \frac{\gamma  \im(\kappa ) \re(\kappa )}{| \re(\kappa )| }.
\end{align}

Thus, the general condition under our assumptions is
\begin{align}
\frac{\gamma  \im(\kappa ) \re(\kappa )}{| \re(\kappa )| }<\gamma  \left(\frac{1}{2}-\im(\Sigma )\right)
\end{align}

Assuming $\re(\kappa) < 0$ and $\im(\kappa)<0$, we can further simplify this to the condition
\begin{align}
\im(\kappa )>\im(\Sigma )-\frac{1}{2}.
\end{align}

\section{Exceptional points} 
In this section we compute the exceptional points of the system and find that they lie far away from the optimal point for sensing in the presented protocol. 
\label{SM:exceptional_points}
The eigenvectors of the effecive Hamiltonian~\eqref{eq:heff} are
\begin{align}
e_{\pm} = \begin{pmatrix} \frac{1}{\tilde \kappa} \left( \Omega_{12} - i \frac{\Gamma_{12}}{2} \pm S\right)\\
1,
\end{pmatrix}
\end{align}
with the eigenvalues given in the main text. 
The energies are degenerate when the square root term $S = \sqrt{- \left( i \Omega_{12} + \frac{\Gamma_{12}}{2}\right)^2 + \tilde \kappa^2  } = 0$. Equating real and imaginary parts, we obtain  
\begin{align}
\gamma_{12} \Gamma_\mathrm{coop}  &= \re\{\tilde \kappa \}\\
\frac{1}{2} \left( \Delta_0 - \Delta\right)  &= \im\{\tilde \kappa\}
\end{align}
and this is the same point at which the eigenvalues also coalesce, i.e., it is the exceptional point for the system.
For any $\gamma_1 \neq \gamma_2$, any degeneracy in the eigenvalue spectrum is an exceptional point.
However, $\gamma_1 = \gamma_2$ is the optimal point for the presented sensing protocol, and the exceptional points lie far away (several orders of magnitude) from this optimum. Therefore, we do not focus on the exceptional points in this work. 
\section{Optimal value of $R$}
\label{SM: optimalR}

By computing $\sigma$ as a function of $R=\gamma_1/\gamma_2$ we found that the optimal value of the gamma ratio is always $R=1$, i.e., that the two impurities have the same decay rate, regardless of the value of $a$, and additionally any value $R=r$ gives the same value of $\sigma$ as its reciprocal $R=1/r$, mirroring the symmetry in the inset to Figure~\ref{fig:impurity_dynamics}b.

\section{Fisher information}
The Fisher information for a binomially distributed quantity is given by $\mathcal{I}(s) = \frac{n}{p(1-p)}$. The variance of a Bernoulli distribution with $n$ trials is $n p(1-p)$, and saturates the Cram\'er-Rao bound for a single shot
\begin{align}
\sigma^2_{c_\mathrm{signal}} \geq \frac{1}{\mathcal{I}(s)} \\ 
n p (1-p) \geq \frac{p(1-p)}{n}.
\end{align}

\section{Propagation of uncertainty for $\sigma_{\Delta_\mathrm{signal}}$}
\label{SM:error_propagation}
Each sampling event is a Bernoulli trial (the population is in the excited state or it is not), where the probability of success is given by $p= |c_1|^2$, corresponding to the probability of finding the detection impurity in the excited state. 

Given we know the uncertainty $\sigma^2_p = p(1-p)$, i.e., it has the well-known variance for a Bernoulli distributed variable, we can propagate the uncertainty through to $\Delta_\mathrm{signal}$, via 

\begin{align}
    \sigma_p^2 &= \left(\frac{\partial p}{\partial \Delta}\right)^2  \sigma_\Delta^2
    \\ \Rightarrow \sigma_{\Delta} &= \frac{\sigma_p}{\left| \frac{\partial p}{\partial \Delta}\right|}.
\end{align} 
Inserting the variance $\sigma^2_p$ yields 
\begin{align}
    \sigma_{\Delta} &= \frac{\sqrt{|c_1|^2(1-|c_1|^2})}{\left| \frac{\partial |c_1|^2}{\partial \Delta}\right|}
\end{align}
Here we note that the derivatives are evaluated at the added detuning $\Delta_\mathrm{add}$.
\section{Calculating the self-energies and effective coupling strengths for non-periodic arrays}
Bloch's theorem can be applied for periodic lattices to obtain the lattice dispersion 
used to determine the self-energies and coupling rates defined in Eqs. \eqref{eqn:self-energy_k} and \eqref{eqn:couplings_k}. For impurities embedded in a non-periodic lattices, the self-energy and coupling rates can be determined via an alternative method based on the real space Hamiltonian.

In the single excitation manifold the system dynamics are governed by the Schrödinger equation $i \partial_t \ket{\psi(t)} = H \ket{\psi(t)}$ with the non-Hermitian Hamiltonian $H = H_{latt} + H_{latt,d} + H_{21} + H_{dd}$, with the individual Hamiltonians defined in Eqs.~\eqref{eqn:H_latt}--\eqref{eqn:H_d_d}. The atomic wave function can be written as $\ket{\psi(t)} = a(t)\ket{G,g_1,g_2} + \sum_{m=1}^{l} b_m(t)e^{i\omega_I t} \ket{e_i,g_1,g_2} + c_1(t) e^{i\omega_I t} \ket{G,e_1,g_2} + c_2(t) e^{i\omega_I t} \ket{G,g_1,e_2}$, where $\ket{G,g_1,g_2}$ denotes the state with all atoms in the ground state, $\ket{e_i,g_1,g_2}$  the state where only the $i$th lattice atom is excited and $\ket{G,e_1,g_2}$ ($\ket{G,g_1,e_2}$) is the state where only the impurity $1$ ($2$) is excited. In the frame rotating at the impurity resonance frequency $\omega_I$, this results in a set of coupled equations for the amplitudes $b_m(t)$, $c_1(t)$ and $c_2(t)$,
\begin{subequations}
\begin{align}
    \partial_t b_m(t) &= i \left(\delta_{LI} + \frac{i}{2} \gamma_L\right) b_m(t)- i \gamma_L\sum_{n\neq m}^l \left(J_{mn}-\frac{i}{2}\Gamma_{mn}\right)b_n(t)\nonumber\\
    & \quad - i \sqrt{\gamma_L\gamma_1}\left(J_{m1}-\frac{i}{2}\Gamma_{m1}\right)c_1(t) - i \sqrt{\gamma_L\gamma_2}\left(J_{m2}-\frac{i}{2}\Gamma_{m2}\right)c_2(t),\label{eqn:bi_eq}\\
    \partial_t c_1(t) &= - \frac{\gamma_1 }{2} c_1(t) -i \sqrt{\gamma_L\gamma_1}\sum_{m=1}^l \left(J_{m1}-\frac{i}{2}\Gamma_{m1}\right)b_i(t) - i\sqrt{\gamma_1\gamma_2}\left(J_{12} - i \frac{\Gamma_{12}}{2}\right)c_2(t)\\
    \partial_t c_2(t) &= - \frac{\gamma_2 }{2} c_2(t) -i \sqrt{\gamma_L\gamma_1}\sum_{m=1}^l \left(J_{m2}-\frac{i}{2}\Gamma_{m2}\right)b_i(t) - i\sqrt{\gamma_1\gamma_2}\left(J_{21} - i \frac{\Gamma_{21}}{2}\right)c_1(t).
\end{align}
\end{subequations}
We introduced the detuning $\delta_{LI} \coloneqq \omega_I - \omega_L$ between the lattice and impurity atom transition frequencies. Note that the model does not contain any classical driving terms so the derivatives of the excited state populations don't depend on the ground state population $a(t)$. This set of equations can be written in matrix form
\begin{equation} i
    \begin{pmatrix}
    \dot{b}_1(t) \\ \vdots \\ \dot{b}_{l}(t) \\ \dot{c_1}(t)\\ \dot{c_2}(t)
    \end{pmatrix} = \begin{pmatrix}
        ~ & ~ & ~ &   c_{1,1} & c_{1,2} \\ 
        ~ & \mathbf{H}_L & ~ & \vdots & \vdots \\
        ~ & ~ & ~ & c_{l,1} & c_{l,2} &\\
        c_{1,1} & \cdots & c_{1, l} & i\gamma_1/2 & c_{1,2}\\
        c_{2,1} & \cdots & c_{2, l} & c_{2,1} & i\gamma_2/2
    \end{pmatrix}
    \cdot
    \begin{pmatrix}
        b_1(t) \\ \vdots \\ b_{l}(t) \\ c_1(t)\\ c_2(t)
    \end{pmatrix} 
    \label{eqn:Seq_matrix}
\end{equation}
where the $l\times l$ matrix $\mathbf{H}_L$ represents the bare lattice Hamiltonian matrix containing the terms $\propto(\delta_{LI} -i\gamma_L/2)$ in the diagonal and the coupling terms $\propto \left(J_{mn} - i\Gamma_{mn}/2\right)$ in the off-diagonals [see first line in~\eref{eqn:bi_eq}]. The complex numbers $c_{i,1} = c_{1,i}$  and $c_{i,2} = c_{2,i}$ with $i\in[1,l]$ denote the coupling terms between the lattice atoms and the impurity $\propto J_{m1} - i\Gamma_{m1}/2$  and $\propto J_{m2} - i\Gamma_{m2}/2$, and $c_{1,2}$ = $c_{2,1}$ denotes the free space coupling of the two impurities $1$ and $2$.

If $\gamma_{1,2} \ll \gamma_L$, the lattice dynamics can be adiabatically eliminated. Defining the quantities $\mathbf{b}(t)\coloneqq (b_1(t) \hdots b_{l}(t))^T$ and the lattice-impurity coupling vectors $\mathbf{C}_{L1}\coloneqq (c_{1,1} \hdots c_{l ,1})^T$, $\mathbf{C}_{L2}\coloneqq (c_{1,2} \hdots c_{l,2})^T$
and setting $\dot{b}_i(t) = 0$ results in the steady state for the lattice atoms,
\begin{equation}
\mathbf{b}_{ss}(t) = - \mathbf{H}_L^{-1} \cdot \left(\mathbf{C}_{L1} c_1(t) + \mathbf{C}_{L2} c_2(t)\right).
\end{equation}
Plugging this back into~\eref{eqn:Seq_matrix}, we obtain the coupled set of equation governing the impurities' dynamics
\begin{align}
    \dot{c_1}(t) &= -i \left[\frac{i}{2}\gamma_1 - \mathbf{C}_{1L}^T \cdot \mathbf{H}_L^{-1}\cdot\mathbf{C}_{L1}\right] c_1(t) - i\left[c_{1,2} - \mathbf{C}_{1L}^T\cdot \mathbf{H}^{-1} \mathbf{C}_{L2}\right]c_2(t),
    \label{eqn:c_dynamics}\\
    \dot{c_2}(t) &= -i \left[\frac{i}{2}\gamma_2 - \mathbf{C}_{2L}^T \cdot \mathbf{H}_L^{-1}\cdot\mathbf{C}_{L2}\right] c_2(t) - i\left[c_{2,1} - \mathbf{C}_{2L}^T\cdot \mathbf{H}^{-1} \mathbf{C}_{L1}\right]c_1(t),
\end{align}
with $\mathbf{C}_{1L}^T \coloneqq (c_{1,1},\hdots,c_{1,l})$ and $\mathbf{C}_{2L}^T \coloneqq (c_{2, 1},\hdots,c_{2, l})$. To obtain the effective Hamiltonian given in~\eref{eqn:Heff} we have to bring these equations into the form
\begin{align}
\dot{c_1}(t) = -i\left[\gamma_1\Sigma_1 - i\frac{\gamma_1}{2}\right]c_1(t) - i\sqrt{\gamma_1\gamma_2}\kappa_1 c_2(t),\\
\dot{c_2}(t) = -i\left[\gamma_2\Sigma_2 - i\frac{\gamma_2}{2}\right]c_2(t) - i\sqrt{\gamma_1\gamma_2}\kappa_2 c_1(t).
\end{align}
This results in the following expressions for the self energies and coupling strengths
\begin{align}
\Sigma_1 \coloneqq - \mathbf{C}_{1L}^T \cdot \mathbf{H}_L^{-1} \cdot \mathbf{C}_{L1}/\gamma_1,\\
\Sigma_2 \coloneqq - \mathbf{C}_{2L}^T \cdot \mathbf{H}_L^{-1} \cdot \mathbf{C}_{L2}/\gamma_2,\\
\kappa_1 \coloneqq \left(c_{1,2} - \mathbf{C}_{2L}^T \cdot \mathbf{H}_L^{-1} \cdot \mathbf{C}_{L1}\right)/\sqrt{\gamma_1\gamma_2},\\
\kappa_2 \coloneqq \left(c_{2,1} - \mathbf{C}_{1L}^T \cdot \mathbf{H}_L^{-1} \cdot \mathbf{C}_{L}\right)/\sqrt{\gamma_1\gamma_2}.
\end{align}

\section{Robustness to noise}
\begin{figure}[h]
    \centering
    \includegraphics[width=0.9\textwidth]{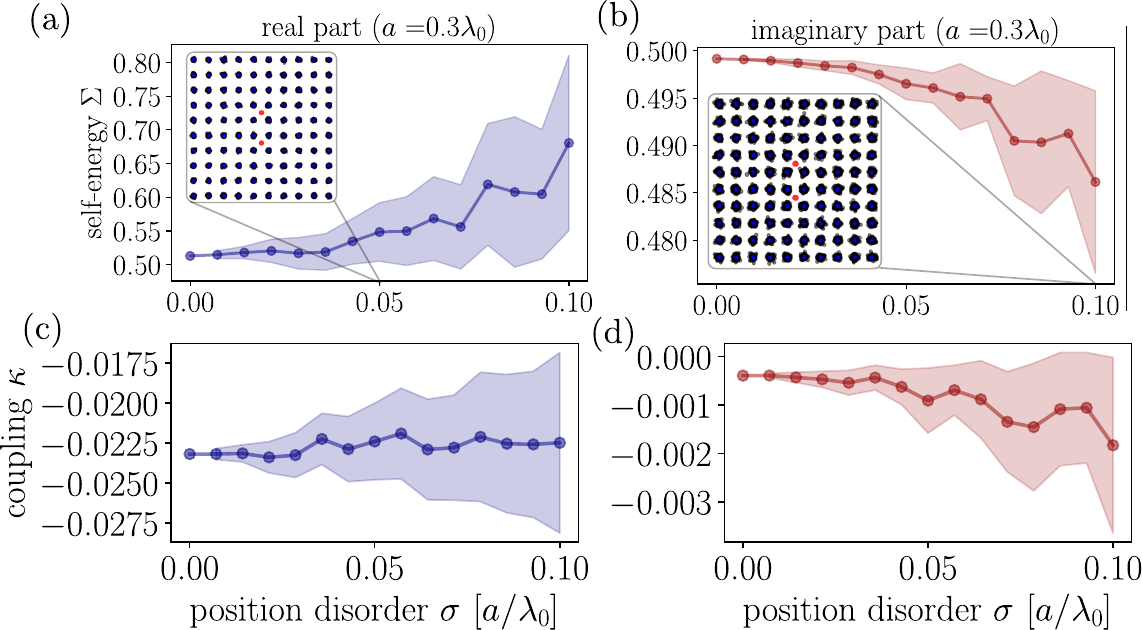}
    \caption{Mean-value and standard deviation (indicated by blue and red shaded regions) of the real- and imaginary-part of the self-energy $\Sigma$ (a)-(b) and the coupling strength $\kappa$ (c)-(d) as a function of position disorder. The insets show the lattice positions realized for a certain position disorders.}
    \label{fig:noise_robustnes}
\end{figure}

 To illustrate the robustness of the introduced sensing protocol to external noise, we analyze the role of positional disorder of the array atoms onto the self-energy and coupling strength.
Therefore, we randomly sample the lattice positions for each emitter from a Gaussian distribution with a certain width $\sigma$ centered around the unperturbed lattice points. We determine $\Sigma$ and $\kappa$ for 100 lattice realizations (for $R=\gamma_1/\gamma_2 = 1$) and plot its mean value together with the standard deviation in~\fref{fig:noise_robustnes}. Note that positional disorder breaks the lattice symmetry and Bloch's theorem, which was employed in the momentum space representation described above, is no longer valid. Therefore, we perform the adiabatic elimination in real space as outlined above.

\begin{figure}
    \centering
    \includegraphics[width=0.9\textwidth]{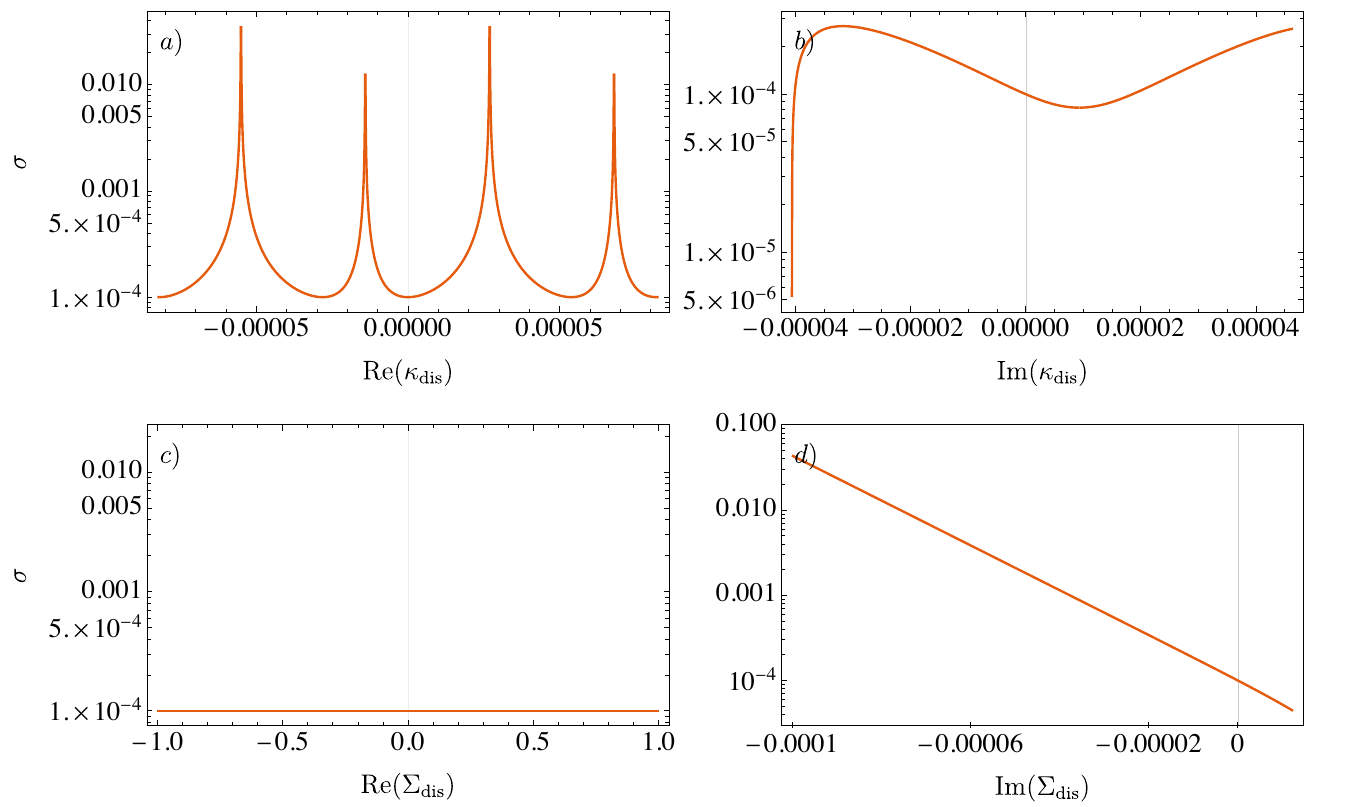}
    \caption{The protocol sensitivity $\sigma$ with the inclusion of lattice disorder. 
    a) $\sigma$ with added real part of the coupling constant $\kappa_\mathrm{dis}$. b) Adding an imaginary $\kappa_\mathrm{dis}$. c) Adding a real part $\Sigma_\mathrm{dis}$ to the self-energy $\Sigma$. c) Adding an imaginary part $\Sigma_\mathrm{dis}$ to the self-energy $\Sigma$.}
    \label{fig:sigma_robustnes}
\end{figure}

The self-energy and the coupling strength are the two quantities at the core of the introduced protocol.
The real part of the coupling constant $\kappa$ changes the time-scale of the dynamics, i.e., it has a similar effect to changing the lattice spacing $a$.
As such, in Fig.~\ref{fig:sigma_robustnes}a), we see a similar response to $\re(\kappa_\mathrm{dis})$ as changing $a$, including crossing over resonance points. 
The imaginary part of $\kappa$ affects the decay lifetime in a similar way to $\Gamma_\mathrm{coop}$. Increasing this parameter can actually \emph{improve} the sensitivity, as seen in 
As discussed in SM~\ref{SM:constraints}, the real part of the self-energy $\Sigma$ simply sets an absolute energy scale and does not affect the dynamics. 
As such, we do not see any change to $\sigma$ in panel c) of Fig.~\ref{fig:sigma_robustnes}.
Finally, altering $\im(\Sigma)$ directly affects $\Gamma_\mathrm{coop}$ and thus can dramatically affect the sensitivity of the protocol. 
In Fig.~\ref{fig:noise_robustnes}b), we see that relatively large amount of lattice disorder is required before a significant change in $\im(\Sigma)$ is observed. 
In panel d) of Fig.~\ref{fig:sigma_robustnes}, we see that the sensitivity remains relatively low for small changes to $\Sigma_\mathrm{dis}$.
However, directly changing the imaginary part of $\Sigma$ is equivalent to changing the cooperative enhancement factor $\Gamma_\mathrm{coop}$ directly. Since improvements to this factor exponentially improve the sensitivity, disorder that destroys this enhancement will similarly see a loss of this exponential improvement. Hence we can see it is important that there is not too much disorder that produces a large change in $\im(\Sigma)$.

\end{document}